# Energy Conversion in Lifting Mass Vertically using a DC Electric Motor by Observing Required Time to Lift Object for a Certain Height

S. Viridi[1,*], A. Purqon[1], S. Permana[1], W. Srigutomo[1], A. Susilawati[2], B. W. Nuryadin[3], Nurhasan[1]

[1]Prodi Fisika, Institut Teknologi Bandung, Jalan Ganesha 10, Bandung 40132, Indonesia
[2]Prodi Fisika, Universitas Padjajaran, Jalan Raya Bandung-Sumedang km 21, Jatinagor 45363, Indonesia
[3]Prodi Fisika, UIN Sunan Gunung Djati, Jalan Raya Cipadung 105, Bandung 10614, Indonesia
[*]dudung@fi.itb.ac.id

**Abstract**. In lifting mass vertically using a DC electric motor energy conversion from electric energy, through intermediate kinetic energy, to gravitation potential energy shows that time required $\Delta t$ to lift load mass $m$ for height $h$ is dependent quadratically to $m$. Several approaches to explain the experiment observation are discussed in this work, from ideal energy conversion to numerical solution from differential equation.

## Introduction

Equipping students with knowledge related to energy education is already a today requirement, especially for living in a world faced with rising energy demands and sinking energy resources (Duit, 1984). It can also sustain students live in the future (Warburton, 2003) and teach them energy transfer technology (Smestad, 1998). Energy as a conserved quantity is more difficult concept for students than its conversion concept (Duit, 2007), since it is an abstract theoretical concept (Lijnse, 1990). While DC motors are already well-known in students daily live in the form of toys, it can simply show conversion energy from chemical potential energy (battery) to kinetic energy (motion), but it still has interesting feature in the torque-speed and power-speed curves (Page, 1999). Inefficiency during energy conversion will be discussed in this work for case lifting mass vertically, where chemical potential energy (battery) is converted to kinetic energy (motor rotation) and then later converted to gravitation potential energy (height change). Several approaches are presented to explain experiment observations

## Theory

Assume that a DC electric motor has resistance $R$. As it is supplied with a voltage difference $V$, then according to Ohm's law it will have current $I$

$$I = \frac{V}{R} \qquad (1)$$

flowing through it. The dissipated power $P_{\mathrm{dis}}$ will be

$$P_{\mathrm{dis}} = VI = \frac{V^2}{R}. \qquad (2)$$

If the motor works for time duration $\Delta t$ then dissipated work can be calculated using

$$W_{\mathrm{dis}} = P_{\mathrm{dis}} \Delta t = \frac{V^2 \Delta t}{R}. \qquad (3)$$

In this system a load with mass $m$ is pulled by the motor and it has height difference $h$. Due to earth gravitation acceleration $g$ the load will has potential energy difference

$$U = mgh. \qquad (4)$$

*Ideal energy conversion*

Assuming that energy conversion from electric energy to kinetic energy and later to gravitation potential energy is totally efficient, then

$$W_{\mathrm{dis}} = U. \qquad (5)$$

Equation (5) will give the relation

$$\Delta t = \left(\frac{ghR}{V^2}\right) m, \qquad (6)$$

which shows linear relation between load pulling duration time $\Delta t$ and load mass $m$.

*Simple DC electric motor configuration*

A direct current (DC) electric motor can be simply modeled as given in Fig. 1. It has a source of permanent magnet with U-form, a rectangle coil, a commutator ring, and a rotor beam. As current $I$ is flowing from positive terminal to negative terminal of the motor, some parts could have non-zero magnetic force resultant, which later leads to magnetic torque, that rotates motor rotor beam.

Supposed that there are $N$ turns of the coil with length $l$ and width $w$, then the magnetic force $F_B$ can be formulated as

$$F_B = NBIl \cos\theta, \qquad (7)$$

where $\theta$ is the angle between area constructed by $l \times w$ and direction of magnetic field $B$. Magnetic torque $\tau_B$ acted on the coil is

$$\tau_B = 2 \cdot \left(\tfrac{1}{2} w \cdot F_B\right) = w F_B. \qquad (8)$$

Substitution Eq. (7) into Eq. (8) will produce

$$\tau_B = NBIlw\cos\theta = NBIA\cos\theta, \qquad (9)$$

where $A$ is area of the coil or simply $l \times w$. Figure 1 shows a condition where $\theta$ is $\pi/2$. In that figure only coil with one turn is shown.

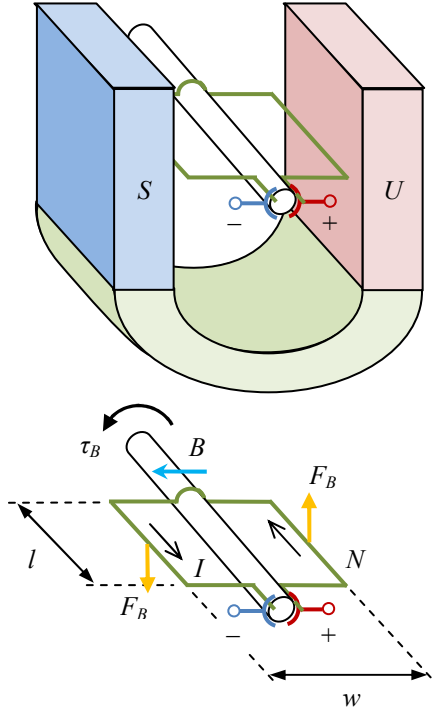

Figure 1. Illustration of a DC electric motor (top) and the force diagram worked on the coil (bottom).

*Load torque*

Load mass $m$ can be attached to motor rotor beam using massless string. If the rotor beam has diameter $D$ then the load mass will introduce load torque

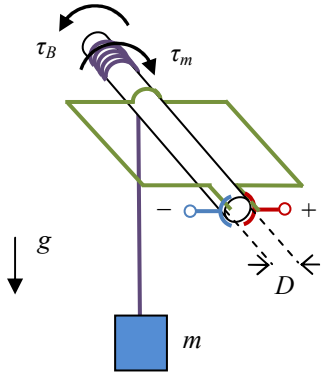

Figure 2. Load mass $m$ introduces load torque $\tau_m$ to motor rotor beam.

$$\tau_m = \tfrac{1}{2}Dm. \qquad (10)$$

*Energy conversion through rotor beam rotation*

Supposed that all rotating parts of the DC motor has moment inertia $I_D$, then

$$\tau_B - \tau_m = I_D\alpha, \qquad (11)$$

is the dinamics of rotor beam. Substitute Eqs. (9), and (10) into Eq. (11), and then also use also Eq. (1) in the result, will give

$$\frac{d^2\theta}{dt^2} - \left(\frac{NBAV}{I_D R}\right)\cos\theta + \frac{Dm}{2I_D} = 0, \qquad (12)$$

a differential equation that specifies rotation of the rotor beam. In order to obtain similar expression as in Eq. (6), Eq. (12) must be solved and

$$\Delta t = \frac{2h}{\theta(\Delta t)D}, \qquad (13)$$

where $\theta(\Delta t)$ is angular position at $t = \Delta t$. Equation (12) will be solved numerically as explained the next section of this work.

*Motor with inductive coil*

Beside that the motor has resistance $R$, it can also be assumed that there is also additional inductive property $L$. This $L$ will play important role while power supply is turned on and current starts to flow to the motor coil. Voltage difference in the coil will be

$$V_L = -L\frac{dI}{dt} \qquad (14)$$

and the same time it has a resistance $R$ in series configuration with $L$, which is similar to Eq. (1)

$$V_R = RI. \qquad (15)$$

Another circuit component that must also be considered it voltage of power supply $V$. According to the second Kirchhoff's law

$$RI - L\frac{dI}{dt} - V = 0 \qquad (16)$$

Equation (9) can be solved

$$I = \frac{V}{R}\left[1 - e^{-(R/L)t}\right] \qquad (17)$$

Equation (17) is quite different than Eq. (1), which is stationer or a condition for Eq. (17) for $t \to \infty$. This equation will modify Eq. (12) to more complex form

$$\frac{d^2\theta}{dt^2} - \left(\frac{NBAV}{I_D R}\right)\left[1 - e^{-(R/L)t}\right]\cos\theta + \frac{Dm}{2I_D} = 0, \quad (18)$$

which will be also solved numerically as Eq. (12). Solution of Eq. (18) will be also used to obtain required time $\Delta t$ to lift the load mass $m$ using Eq. (13).

*Empirical model*

A function of required time $\Delta t$ as function of load mass $m$ is simply defined as

$$\Delta t = c_2 m^2 + c_1 m + c_0, \qquad (19)$$

where the constants $c_2$, $c_1$, and $c_0$ will be obtained by fitting experiment data with Eq. (19).

**Numerical Method**

Euler method is used to solve Eqs. (12) and (18). The method relates a variable and its time derivative through

$$f(t + \Delta t) = f(t) + \Delta t \frac{df(t)}{dt}. \qquad (20)$$

Equation (12) can be written in following equations

$$\frac{d^2\theta(t)}{dt^2} = \left(\frac{NBAV}{I_D R}\right)\cos\theta(t) - \frac{Dm}{2I_D}, \quad (21)$$

$$\frac{d\theta(t+\Delta t)}{dt} = \frac{d\theta(t)}{dt} + \Delta t \frac{d^2\theta(t)}{dt^2}, \quad (22)$$

$$\theta(t+\Delta t) = \theta(t) + \Delta t \frac{d\theta(t+\Delta t)}{dt}. \quad (23)$$

For Eq. (18), modification must be applied to Eq. (21)

$$\frac{d^2\theta(t)}{dt^2} = \left(\frac{NBAV}{I_D R}\right)\left[1 - e^{-(R/L)t}\right]\cos\theta(t) - \frac{Dm}{2I_D}, \quad (24)$$

where Eqs. (22) and (23) can still be used.

### Experiment

Setup of the experiment is given in Fig. 3, which is similar to task 4 while probing only the concept (Duit, 1984).

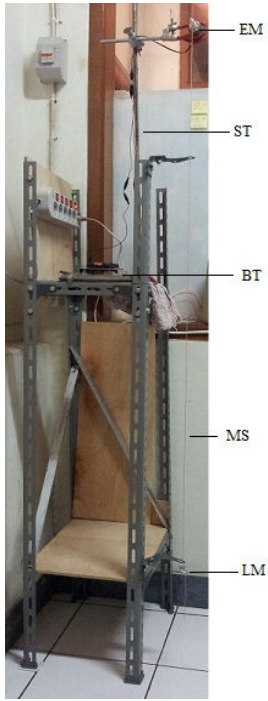

Figure 3. Experiment setup: DC electric motor (EM), statif (ST), battery (BT), massless string (MS), and load mass (LM).

To simplify mass measurement pairs of nut and bolt are used (Fig. 4) and their number is the unit of mass instead of using regular unit such as kg or g. It is hoped that student attention in observing energy converstion is not distracted to the mass of measurement and they can also learn that mass can be quantisized in arbitrary unit as long as it is known.

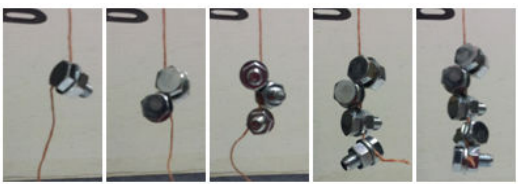

Figure 4. Load mass quantization using pairs of nut and bold.

Change in vertical position is set to be constant $h = 1$ m. Battery voltage difference is varied for 3, 6, and 9 V. Load mass is also varied for 1-5 pieces (pairs of nut and bolt). In each variation time required to lift the mass is observed five times, averaged, and reported as $\Delta t$.

### Results and Discussion

In experiment load mass is quantized using nut and bolt (piece) instead of kg or g. Observation of required time $\Delta t$ to lift the mass $m$ fo height $h$ is given in Fig. 5.

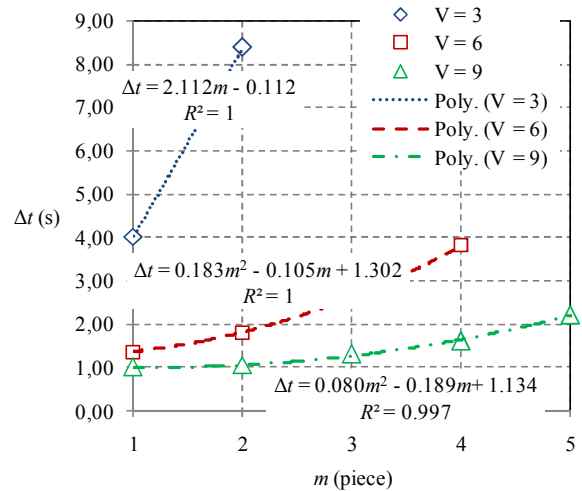

Figure 5. Experiments observation of required time $\Delta t$ to lift the load mass $m$ for height $h$.

Sinc Eq. (6) is a linear function of $m$, it will only fit for $V = 3$ V in experiment observation and fails for other value of $V$. From the equation electric motor coil resistance is found

$$\left(\frac{ghR}{V^2}\right) = 2.112 \Rightarrow R = \frac{2.112 \cdot 3^2}{9.81 \cdot 1} = 1.938 \; \Omega.$$

### Conclusion

Model to calculate required time in lifting load mass vertically using a DC motor for certain height change has been obtained empirically, which is a second order polynomial function of load mass. Unfortunately, theoretical approaches using ideal energy conversion and through rotation dynamics have failed to explain experiment observation.

### Acknowledgements

SV would like to than RIK-ITB in year 2014 for partially supporting this work and Kemenag Republik Indonesia for trigerring idea for this work.